\documentclass[11pt]{article}
\usepackage{amsfonts,amsmath,amssymb}
\usepackage{enumerate}
\usepackage{hyperref}
\usepackage{bbm}
\usepackage{nicefrac}
\usepackage[all]{xy}
\usepackage{graphicx}
\usepackage{bm}
\usepackage{makecell}
\usepackage[table]{xcolor}
\usepackage{longtable}		
\usepackage{subcaption}		
\usepackage{tikz}			

\usepackage{upgreek}

\usepackage{float}			
\usepackage{lscape}                 

\usepackage{booktabs}

\newcommand{\be}{\begin{equation}}
\newcommand{\ee}{\end{equation}}

\newcommand{\bea}{\begin{eqnarray}}
\newcommand{\eea}{\end{eqnarray}}

\newcommand{\bes}{\begin{subequations}}
\newcommand{\ees}{\end{subequations}}

\newcommand{\cN}{{\cal N}}

\def\sst#1{{\scriptscriptstyle #1}}

\def\0{{\sst{(0)}}}
\def\1{{\sst{(1)}}}
\def\2{{\sst{(2)}}}
\def\3{{\sst{(3)}}}
\def\4{{\sst{(4)}}}
\def\5{{\sst{(5)}}}
\def\6{{\sst{(6)}}}
\def\7{{\sst{(7)}}}
\def\8{{\sst{(8)}}}

\def\cM{{{\cal M}}}

\allowdisplaybreaks  

\usepackage{multirow}
\usepackage{rotating}

\newcommand{\ba}{\begin{align}}
\newcommand{\ea}{\end{align}}

\newcommand{\bse}{\begin{subequations}}
\newcommand{\ese}{\end{subequations}}

\allowdisplaybreaks

\usetikzlibrary{arrows}

\newcommand{\mon}{
	\begin{tikzpicture}
		\draw[->, >=stealth', semithick](0,0)-- (35:15pt);
	\end{tikzpicture}
}

\newcommand{\nonmon}{
	\begin{tikzpicture}[->, >=stealth', semithick]		
		\node[] at (0,0) (1) {};
		\node[] at (1,0.25) (2) {};
		\draw (1) [out=-30, in=-140] to  (2);
	\end{tikzpicture}
}

\newcommand{\uA}{\underline{A}}
\newcommand{\uB}{\underline{B}}
\newcommand{\uC}{\underline{C}}
\newcommand{\uD}{\underline{D}}
\newcommand{\uE}{\underline{E}}
\newcommand{\uF}{\underline{F}}
\newcommand{\uG}{\underline{G}}

\usepackage{lipsum}

\newlength\Colsep
\begin{document}

\makeatletter
\renewcommand{\theequation}{\thesection.\arabic{equation}}
\@addtoreset{equation}{section}
\makeatother

\begin{titlepage}

\begin{flushright}
IFT-UAM/CSIC-21-025 \\
%
\end{flushright}

\vspace{5pt}

   \begin{center}
   \baselineskip=16pt

   \begin{Large}\textbf{
\hspace{-18pt} Supersymmetric spectroscopy on $\textrm{AdS}_4 \times S^{ 7}$   and $\textrm{AdS}_4 \times S^{ 6}$   \\[8pt]
}
   \end{Large}

\vspace{25pt}

{\large  Mattia Ces\`aro$^{1}$, Gabriel Larios$^{1}$ \,and \,  Oscar Varela$^{1,2}$}
		
\vspace{25pt}

	\begin{small}

	{\it $^{1}$ Departamento de F\'\i sica Te\'orica and Instituto de F\'\i sica Te\'orica UAM/CSIC , \\
   Universidad Aut\'onoma de Madrid, Cantoblanco, 28049 Madrid, Spain}  \\

	\vspace{15pt}
	
	{\it $^{2}$ Department of Physics, Utah State University, Logan, UT 84322, USA}     \\	
		
	\end{small}

\vskip 50pt

\end{center}

\begin{center}
\textbf{Abstract}
\end{center}

\begin{quote}

New techniques based on Exceptional Field Theory have recently allowed for the calculation of the Kaluza-Klein spectra of certain AdS$_4$ solutions of $D=11$ and massive IIA supergravity. These are the solutions that consistently uplift on $S^7$ and $S^6$ from vacua of maximal four-dimensional supergravity with SO(8) and ISO(7) gaugings. In this paper, we provide an algorithmic procedure to compute the complete Kaluza-Klein spectrum of five such AdS$_4$ solutions, all of them ${\cal N}=1$, and give the first few Kaluza-Klein levels. These solutions preserve SO(3) and $\textrm{U}(1) \times \textrm{U}(1)$ internal symmetry in $D=11$, and U(1) (two of them) and no continuous symmetry in type IIA. Together with previously discussed cases, our results exhaust the Kaluza-Klein spectra of known supersymmetric AdS$_4$ solutions in $D=11$ and type IIA in the relevant class.

\end{quote}

\vfill

\end{titlepage}

\tableofcontents



\section{Introduction}


The spectrum of Kaluza-Klein (KK) perturbations above backgrounds of string and M-theory with lower-dimensional anti-de Sitter factors carries important physical information about such solutions. The complete KK spectrum of some supersymmetric solutions of $D=11$ supergravity \cite{Cremmer:1978km} of Freund-Rubin (FR) type \cite{Freund:1980xh} has long been known. Recall that FR solutions involve the direct product, AdS$_4 \times M_7$, of four-dimensional anti-de Sitter space (AdS$_4$) with a seven-dimensional Einstein manifold $M_7$, supported by four-form flux on AdS$_4$, and may preserve various amounts $\cN$ of supersymmetry \cite{Acharya:1998db}. A maximally supersymmetric, $\cN=8$, FR solution is obtained when $M_7$ is chosen to be the seven-sphere, $S^7$. The KK spectrum of this solution was first determined using group theory arguments based on the fact that, in this case, all states belong to short multiplets of the supersymmetry algebra OSp$(4|8)$ \cite{Englert:1983rn,Sezgin:1983ik,Biran:1983iy}. In contrast, the KK spectra of FR solutions with $1 \leq \cN < 8$ typically contain both short and long multiplets of OSp$(4|\cN)$. Thus, group theory does not suffice to determine the KK spectra in those cases, and direct calculation is necessary. For the subclass of FR solutions where $M_7$ is homogeneous, specific coset-space techniques have been used to compute the complete KK spectra of particular $\cN=3$ \cite{Fre:1999gok}, $\cN=2$ \cite{Fabbri:1999mk} and $\cN=1$ \cite{Nilsson:2018lof} solutions.

The FR class represents only a small sector of AdS$_4$ solutions of $D=11$ supergravity. More general $1 \leq {\cal N} <8$ solutions are known that involve warped products, AdS$_4 \times_w M_7$, with necessarily inhomogeneous \cite{Cassani:2012pj} non-Einstein metrics on $M_7$ and threaded by internal fluxes. See for example \cite{Lukas:2004ip,Behrndt:2005im,Gabella:2012rc} for some classification results in $D=11$ \cite{Cremmer:1978km}, and \cite{Behrndt:2004km,Lust:2004ig,Passias:2018zlm} in massive type IIA \cite{ Romans:1985tz}, the $D=10$ supergravity that will be relevant in this paper together with $D=11$. Typically, group theory is again enough to determine the KK spectrum of short OSp$(4|\cN)$ multiplets, see \cite{Klebanov:2008vq} for an example. But the complete spectrum will usually contain long multiplets as well, whose dimensions can only be obtained by direct calculation. This computation can still be relatively easily carried out for the graviton sector of the KK spectrum, see {\it e.g.} \cite{Klebanov:2009kp,Richard:2014qsa,Pang:2015rwd,Pang:2017omp,Dimmitt:2019qla,Cesaro:2020piw}, using the specific spin-2 methods of \cite{Bachas:2011xa}. But, in general, the calculation of the complete KK spectrum in the vector, scalar and fermion sectors will necessarily involve an implementation of the standard techniques of \cite{Kim:1985ez}. On such complicated warped, flux, backgrounds, the calculation following \cite{Kim:1985ez} quickly becomes extremely hard and, for that reason, progress on finding the  KK spectra of such solutions has been virtually inexistent over the last twenty years or so.

This situation has recently changed. An alternative to the methods of \cite{Kim:1985ez} has been proposed \cite{Malek:2019eaz} based on Exceptional Field Theory (ExFT) \cite{Hohm:2013pua,Hohm:2013uia,Godazgar:2014nqa,Ciceri:2016dmd}, a duality-covariant reformulation (see also \cite{Pacheco:2008ps,Berman:2010is,Cassani:2016ncu} and \cite{Berman:2020tqn} for a review) of the higher-dimensional supergravities. While these novel ExFT techniques are not universally valid for all solutions in the generic classifications of {\it e.g.} \cite{Lukas:2004ip,Behrndt:2005im,Gabella:2012rc,Behrndt:2004km,Lust:2004ig,Passias:2018zlm}, they nonetheless hold for a very interesting subclass of solutions: those, supersymmetric or otherwise, that consistently uplift on topological spheres from lower-dimensional gauged maximal supergravities. In this paper we will focus on $D=11$  \cite{Cremmer:1978km}  and massive type IIA \cite{Romans:1985tz} supersymmetric solutions of the form AdS$_4 \times_w S^7$ and AdS$_4 \times_w S^6$, that uplift consistently \cite{deWit:1986iy,Guarino:2015jca,Guarino:2015vca} from critical points of $D=4$ $\cN=8$ supergravity \cite{deWit:2007mt} with SO(8) \cite{deWit:1982ig} and dyonic ISO(7) \cite{Guarino:2015qaa} gaugings, respectively. See \cite{Comsa:2019rcz,Bobev:2020qev} for the most recent surveys of this class of supersymmetric and non-supersymmetric solutions, and table \ref{tab:AllSusyAdS4} for a summary of all known supersymmetric solutions in this class. We will diagonalise the mass matrices for the KK perturbations above this type of solutions as given in \cite{Malek:2019eaz,Dimmitt:2019qla,Varela:2020wty,Cesaro:2020soq}. For convenience, we review these mass matrices in section \ref{sec:KKMassMatRev}.

A powerful feature of these ExFT-based techniques is that the resulting KK mass matrices only depend on data of the relevant $D=4$ $\cN=8$ gauged supergravity along with the generators of SO(8) or SO(7). The latter may be seen as remnants of the internal, uplifting topological spheres $S^7$ or $S^6$. In any case, unlike the standard methods of \cite{Kim:1985ez} or even the spin-2 techniques of \cite{Bachas:2011xa}, knowledge of the full higher-dimensional solutions is not needed to compute their complete KK spectra. The solutions need only be known as critical points of the corresponding $D=4$ $\cN=8$ gauged supergravity. In fact, none of the AdS$_4$ solutions we compute the KK spectrum of in this paper has actually been constructed in fully-fledged ten- or eleven-dimensional form. More concretely, in this work we present the KK spectrum of the AdS$_4 \times_w S^7$ solutions of $D=11$ supergravity that uplift from the $\cN=1$ critical points of $D=4$ $\cN=8$ SO(8) supergravity \cite{deWit:1982ig} with SO(3) \cite{Comsa:2019rcz,Bobev:2019dik} and $\textrm{U}(1) \times \textrm{U}(1)$ \cite{Fischbacher:2009cj,Fischbacher:2010ec} residual symmetry. We also provide the KK spectra of the AdS$_4 \times_w S^6$ solutions of massive IIA supergravity that uplift from the $\cN=1$ vacua of $D=4$ $\cN=8$ ISO(7) supergravity \cite{Guarino:2015qaa} with U(1) (two of these) \cite{Guarino:2019snw} and no leftover continuous symmetry \cite{Bobev:2020qev}.

Together with the complete KK spectra of the $D=11$ and type IIA supersymmetric AdS$_4$ solutions summarised in table \ref{tab:AllSusyAdS4}, that have been previously computed in \cite{Englert:1983rn,Sezgin:1983ik,Biran:1983iy,Klebanov:2008vq,Malek:2020yue,Varela:2020wty,Cesaro:2020soq}, our results exhaust the spectra for all known such supersymmetric AdS$_4$ solutions. Sections \ref{sec:newspectra} and \ref{sec:discussion} discuss at length these new KK spectra, the first few KK levels of which have been given in ancillary files for convenience, see appendix \ref{sec:TablesKKSpec}. See \cite{Malek:2020mlk,Guarino:2020flh,Eloy:2020uix,Bobev:2020lsk,Giambrone:2021zvp} for further, possibly partial, results on the KK spectrum of other AdS solutions of string/M-theory recently computed using ExFT techniques.

\begin{table}[]


\resizebox{\textwidth}{!}{

\begin{tabular}{cccccc}
\hline
\\[-2.5mm]
susy  	& $G$ &  $g^{-2}   V$ &  $D=4$ & $D=11$ & Kaluza-Klein 
\\[0pt]
          	&               &         &       solution         &  uplift            &  spectrum
\\[2pt]
\hline
\\[-10pt]
$ \cN=8 $ &	$\textrm{SO}(8)$  &  $-24 $   &	\cite{deWit:1982ig} & \cite{Freund:1980xh}  &\cite{Englert:1983rn,Sezgin:1983ik,Biran:1983iy}  
\\[10pt]
$ \cN=2 $ & $\textrm{SU}(3) \times \textrm{U}(1)$  &   $-18\sqrt3 $&	\cite{Warner:1983vz}  & \cite{Corrado:2001nv} & \cite{Klebanov:2008vq,Malek:2020yue} 
\\[10pt]
$ \cN=1 $ &	G$_2$ & $-\frac{2^{\nicefrac{11}{2}}\;3^{\nicefrac{13}{4}}}{5^{\nicefrac{5}{2}}}$   &	 \cite{Warner:1983vz}  & \cite{deWit:1984nz,Godazgar:2013nma} & \cite{Cesaro:2020soq}  
\\[10pt]
$ \cN=1 $ &	SO(3) &   $ -55.363855$  &	\cite{Comsa:2019rcz,Bobev:2019dik}  & N.A.  &  [here]    
\\[10pt]
$ \cN=1 $ &	$\textrm{U}(1) \times \textrm{U}(1)$ &  $-48$  &	\cite{Fischbacher:2009cj,Fischbacher:2010ec}  & N.A.  &  [here]   
\\[10pt]
\phantom{$ \cN=1 $}  &	\phantom{$\textrm{U}(1) \times \textrm{U}(1)$ } & \phantom{  $(...)$  } &	\phantom{ \cite{Fischbacher:2009cj,Fischbacher:2010ec} } & \phantom{ N.A. } & \phantom{ [here, table (...)]   }
\\[10pt]
\phantom{$ \cN=1 $}  &	\phantom{$\textrm{U}(1) \times \textrm{U}(1)$ } & \phantom{  $(...)$  } &	\phantom{ \cite{Fischbacher:2009cj,Fischbacher:2010ec} } & \phantom{ N.A. } & \phantom{ [here, table (...)]   }
\\[10pt]
\hline
\end{tabular}

\quad

\begin{tabular}{cccccc}
\hline
\\[-2.5mm]
susy  	& $G$ &  $g^{-2}  c^{\frac13} V$ &  $D=4$ & IIA & Kaluza-Klein 
\\[0pt]
          	&               &         &       solution         &  uplift            &  spectrum
\\[2pt]
\hline
\\[-10pt]
$ \cN=3 $ &	$\textrm{SO}(3) \times \textrm{SO}(3)$ &  $-\frac{2^{16/3}}{3^{1/2}} $   &	\cite{Gallerati:2014xra} & \cite{Pang:2015vna,DeLuca:2018buk}  &\cite{Varela:2020wty}  
\\[10pt]
$ \cN=2 $ & $\textrm{SU}(3) \times \textrm{U}(1)$  &   $-2^2 \, 3^{3/2} $&	\cite{Guarino:2015jca}  & \cite{Guarino:2015jca} & \cite{Varela:2020wty} 
\\[10pt]
$ \cN=1 $ &	G$_2$ & $- \frac{2^{28/3} \, 3^{1/2}}{5^{5/2}} $   &	\cite{Borghese:2012qm}  & \cite{Behrndt:2004km,Varela:2015uca} & \cite{Cesaro:2020soq}  
\\[10pt]
$ \cN=1 $ &	SU(3) &   $-\frac{2^{8} \, 3^{3/2}}{5^{5/2}} $  &	\cite{Guarino:2015qaa}  & \cite{Varela:2015uca}  & \cite{Cesaro:2020soq}  
\\[10pt]
$ \cN=1 $ &	$\textrm{U}(1)$ &  $-25.697101$  &	\cite{Guarino:2019snw}  & N.A.  &  [here]   
\\[10pt]
$ \cN=1 $ &	$\textrm{U}(1)$ &  $-35.610235$  &  \cite{Guarino:2019snw}  &	N.A. &  [here]   
\\[10pt]
$ \cN=1 $ &	$\emptyset$ &  $-35.598340$  &  \cite{Bobev:2020qev}  &	N.A. &  [here]   
\\[5pt]
\hline
\end{tabular}

}

\caption{\footnotesize{All known supersymmetric AdS$_4$ solutions that consistently uplift to $D=11$ (left) and massive IIA (right) supergravities from $D=4$ $\cN=8$ supergravity with SO(8) and dyonic ISO(7) gaugings, respectively. For every solution it is shown its residual supersymmetry $\cN$, bosonic symmetry $G$, and $D=4$ cosmological constant $V$ in units of the gauge coupling $g$ and dyonic parameter $c$ (if applicable). Pointers are given to the references where the solutions were found within $D=4$ and $D=11$ or IIA (if available), and to their KK spectra.
}\normalsize}
\label{tab:AllSusyAdS4}
\end{table}


\section{Kaluza-Klein mass matrices} \label{sec:KKMassMatRev}


The mass matrices for the bosonic and fermionic KK perturbations above the AdS$_4$ solutions of interest have been derived from E$_{7(7)}$ ExFT \cite{Hohm:2013pua,Hohm:2013uia,Godazgar:2014nqa,Ciceri:2016dmd} in \cite{Malek:2019eaz,Malek:2020yue,Varela:2020wty,Cesaro:2020soq}. See also \cite{Dimmitt:2019qla} for an early derivation of the KK graviton mass matrix. In this section we review these mass matrices for later reference, and make some new comments about the KK vector mass matrix.

Let us start with the KK fermionic sector. The KK gravitino mass matrix is proportional to \cite{Cesaro:2020soq}
\begin{equation} \label{eq:A1KKdd}
 A_{1 \, i \Lambda , j \Sigma} = A_{1 ij} \, \delta_{\Lambda \Sigma} - 8 \, ( {\cal V}^{-1} )_{ij}{}^{ \underline{M} }  ({\cal T}_{\underline{M}})_{\Lambda \Sigma} \; ,
\end{equation}
and the KK spin$-1/2$ fermion mass matrix proportional to \cite{Cesaro:2020soq}
\begin{equation} \label{eq:A3dddddd}
A_{3 \,  ijk \Lambda , lmn \Sigma}  = A_{3 \,  ijk , lmn} \, \delta_{\Lambda \Sigma} + \tfrac{\sqrt{2}}{18} \,  \epsilon_{ijklmnpq} \, {\cal} ( {\cal V}^{-1} )^{pq \underline{N} }  ({\cal T}_{\underline{N}})_{\Lambda \Sigma} \; .
\end{equation}
Indices $\underline{M}$ and $i$ here and elsewhere label the fundamental representations of E$_{7(7)}$ and SU(8), respectively, while  $\Lambda$ runs over the infinite tower formed by the symmetric-traceless representations $[n,0,0,0]$ of SO(8) or $[n,0,0]$ of SO(7), at all KK level $n=0,1,2,\ldots$, for solutions that respectively uplift from the SO(8) or ISO(7) gaugings of maximal supergravity. The latter indices may be raised and lowered with $\delta_{\Lambda\Sigma}$. The quantities $A_{1 ij} $, $A_{3 \,  ijk , lmn}$ and $( {\cal V}^{-1} )_{ij}{}^{ \underline{M} } $ pertain to $D=4$ $\cN=8$ supergravity \cite{deWit:2007mt} and depend on the $\cN=8$ scalar fields and (the former two only) on the embedding tensor. More concretely, they  correspond respectively to the $\cN=8$ gravitino mass matrix, the ${\cal N}=8$ spin$-1/2$ fermion mass matrix, and to (a subset of the components of) the inverse coset representative $( {\cal V}^{-1} )_{\underline{A} }{}^{ \underline{M} } = \left(  ( {\cal V}^{-1} )_{ij }{}^{ \underline{M} } , ( {\cal V}^{-1} )^{ ij \underline{M} } \right) $ of $\textrm{E}_{7(7)}/\textrm{SU}(8)$, with $\underline{A}$ an index in the $\mathbf{28} + \overline{\bm{28}}$ of SU(8). Finally, the constant, antisymmetric matrices $({\cal T}_{\underline{M}})_{\Lambda \Sigma}$ contain the generators of SO(8) or SO(7): see appendix A of \cite{Varela:2020wty} for our conventions.

Moving on into the bosonic sector, the KK graviton mass matrix is simply \cite{Dimmitt:2019qla,Malek:2019eaz} 
\begin{equation} \label{eq:KKGravMassMat}
({\cal M}_\textrm{grav}^2)_{\Lambda \Sigma} = M^{\underline{M} \underline{N}} \, \delta_{\Omega \Omega^\prime} \,  ({\cal T}_{\underline{M}})_\Lambda{}^\Omega \,  ({\cal T}_{\underline{N}})_\Sigma{}^{\Omega^\prime}  \; ,
\end{equation}
while the KK vector mass matrix reads \cite{Varela:2020wty}
{\setlength\arraycolsep{2pt}
\begin{eqnarray} \label{eq:KKVecMassMat}
({\cal M}^2_\textrm{vec})_{\underline{M} \Lambda}{}^{\underline{N} \Sigma} &=& \tfrac{1}{12}   \,  \delta_{\Omega^\prime\Omega^{\prime\prime}} \,  \Big(  X_{\underline{M} \Lambda \,  (\underline{R } | }{}^{\underline{T} \Omega^\prime } X_{\underline{P} \Omega \,  \underline{U} }{}^{(\underline{R } |  \Omega^{\prime\prime} } + X_{\underline{P} \Omega \,  (\underline{R } | }{}^{\underline{T} \Omega^\prime } X_{\underline{M} \Lambda \,  \underline{U} }{}^{(\underline{R } |  \Omega^{\prime\prime} }   \Big)  \nonumber \\[5pt]
&&
\qquad  \qquad \times  M_{| \underline{S}) \underline{T} } \,   M^{|\underline{S}) \underline{U} } \, M^{\underline{P} \underline{N}} \, \delta^{\Omega \Sigma }   \; ,
\end{eqnarray}
}in terms of the quantity
\begin{equation} \label{KKXSymbols}
X_{\underline{M} \Lambda \,  \underline{N} }{}^{\underline{P} \Sigma }  \equiv \Big( X_{\underline{M}   \underline{N} }{}^{\underline{P}  }   \, \delta_\Lambda^{\Sigma} -12 \, \mathbb{P}^{\underline{P}}{}_{\underline{N}}{}^{\underline{Q}}{}_{\underline{M}} \,   \, ({\cal T}_{\underline{Q}})_\Lambda{}^{\Sigma} \Big) \; . 
\end{equation}
In (\ref{eq:KKGravMassMat}) and (\ref{eq:KKVecMassMat}), $M_{\underline{M} \underline{N}} \equiv \eta_{\underline{A} \underline{B}} \, {\cal V}_{\underline{M}}{}^{\underline{A}}\, {\cal V}_{\underline{N}}{}^{\underline{B}}$, and $M^{\underline{M} \underline{N}}$ its inverse, with $\eta_{\underline{A} \underline{B}}$ the SO$(28,28)$ invariant metric. In (\ref{KKXSymbols}), $X_{\underline{M}   \underline{N} }{}^{\underline{P}} $ are the usual $D=4$ $\cN=8$ $X$-symbols codifying the embedding tensor \cite{deWit:2007mt}, and 
$\mathbb{P}^{\underline{P}}{}_{\underline{N}}{}^{\underline{Q}}{}_{\underline{M}} \equiv  (t_\alpha)_{\underline{N}}{}^{\underline{P}}  (t^\alpha)_{\underline{M}}{}^{\underline{Q}}$ is the projector, written in terms of the generators $ (t_\alpha)_{\underline{M}}{}^{\underline{N}}$ of E$_{7(7)}$, onto the adjoint of the latter. 

The KK vector mass matrix (\ref{eq:KKVecMassMat}) is manifestly E$_{7(7)}$ covariant. An alternate form of this vector mass matrix has appeared in \cite{Malek:2020yue} (see (4.31)  therein), written instead with SU(8) indices. The latter may be brought to the form
\begin{equation}\label{eq:KKVecMassMatSU8}
\left(\cM^2_\textrm{vec} \right)_{\uA  \Lambda}{}^{\uB \Sigma} = \tfrac{1}{12} \, T_{\uA \Lambda\; \uC}{}^{\uD \Omega^\prime}   \left( T_{\uG  \Omega\; \uD}{}^{\uC  \Omega^{\prime\prime}}  + T_{\uG \Omega\; \uE}{}^{\uF \Omega^{\prime\prime}} \,  \eta_{\uF\uD}\eta^{\uC \uE} \right)  \delta_{\Omega^\prime \Omega^{\prime\prime}} \, \eta^{\uG \uB} \, \delta^{\Omega \Sigma}  \, ,
\end{equation}
where we have defined
\begin{equation} \label{eq:genTtensor}
T_{\underline{A} \Lambda \; \underline{B}}{}^{\underline{C}  \Sigma} \equiv T_{\uA \uB}{}^{\uC} \,\delta_\Lambda^\Sigma-12 \,\mathbb{P}^{\uC}{}_{\uB}{}^{\uD}{}_{\uA} \,  (\mathcal{T}_{\uD} )_{\Lambda}{}^\Sigma \; ,
\end{equation}
featuring the $T$-tensor $T_{\uA \uB}{}^{\uC} $ of $D=4$ $\cN=8$ supergravity \cite{deWit:2007mt}. The quantity (\ref{eq:genTtensor}) is the SU(8) version of (\ref{KKXSymbols}), obtained through contractions of the latter with the $\textrm{E}_{7(7)}/\textrm{SU}(8)$ coset representative and its inverse, 
\begin{equation}  
T_{\underline{A} \Lambda \; \underline{B}}{}^{\underline{C}  \Sigma}  =  ( {\cal V}^{-1} )_{\underline{A} }{}^{ \underline{M} }  \, ( {\cal V}^{-1} )_{\underline{B} }{}^{ \underline{N} } \, {\cal V}_{\underline{P} }{}^{ \underline{C} }  \, X_{\underline{M} \Lambda \,  \underline{N} }{}^{\underline{P} \Sigma } \; ,
\end{equation}
exactly as for the relation between the $X$- and $T$-tensors of $D=4$ $\cN=8$ supergravity \cite{deWit:2007mt}. This makes apparent the equivalence between the two vector mass matrices (\ref{eq:KKVecMassMat}) and (\ref{eq:KKVecMassMatSU8}), and also the equivalence to (4.31) of \cite{Malek:2020yue}.

On a related note, the vector mass matrix of $D=4$ $\cN=8$ gauged supergravity can be written in terms of the $\cN=8$ fermion shifts $A_{1 ij} $ and $A_{2 \,  h}{}^{ijk}$, see (4.83), (4.84) of \cite{Trigiante:2016mnt}. Similarly, splitting the indices in the $\mathbf{28} + \overline{\bm{28}}$ of SU(8) in terms of fundamental indices as $\uA = \big({}_{[ij]},\, {}^{[ij]}\big)$, the KK vector mass matrix (\ref{eq:KKVecMassMatSU8}) with (\ref{eq:genTtensor}) takes on the block structure
\begin{equation}
\left(\cM^2_\textrm{vec} \right)_{\uA  \Lambda}{}^{\uB \Sigma} = \left(
\begin{array}{cc}
\left(\cM^2_\textrm{vec} \right)_{ij  \Lambda}{}^{lm  \Sigma} & \left(\cM^2_\textrm{vec} \right)^{ij \, \Omega\; lm \, \Sigma} \,  \delta_{\Omega \Lambda} \\
\left(\cM^2_\textrm{vec} \right)_{ij  \Lambda\; lm  \Omega} \delta^{\Omega \Sigma} & \left(\cM^2_\textrm{vec} \right)^{ij  \Omega}{}_{lm  \Omega^\prime} \, \delta^{\Omega^\prime \Sigma} \, \delta_{\Lambda\Omega}
\end{array}
\right)
\end{equation}
with
$\left(\cM^2_\textrm{vec} \right)_{ij \Lambda}{}^{lm  \Omega} = \left( \left(\cM^2_\textrm{vec} \right)^{ij \Lambda}{}_{lm  \Omega}\right)^*$ 
and 
$\left(\cM^2_\textrm{vec} \right)_{ij  \Lambda \;lm \Omega} = \left(\left(\cM^2_\textrm{vec} \right)^{ij  \Lambda \; lm  \Omega}\right)^*$, 
and
{\setlength\arraycolsep{2pt}
\begin{eqnarray} 
\label{eq:BlockMvec1}
 \nonumber\left(\cM^2_\textrm{vec} \right)^{ij  \Lambda}{}_{lm  \Omega}& = &\frac{1}{12} \left(-A_2^{[i}{}_{pqr}\delta^{j]}_{[l}A_{2\; m]}{}^{pqr}+3 A_2^{[i}{}_{pq[l}A_{2\; m]}{}^{j]pq}\right) \delta^\Lambda_\Omega \\[4pt] \nonumber
&& +\Big(\delta^{[i}_{[l}A_{2\; m]}{}^{j]pq}\left(\mathcal{T}_{pq}\right)^{\Lambda}{}_\Omega+A_{2\; [m}{}^{ijq}\left(\mathcal{T}_{l]q}\right)^{\Lambda}{}_\Omega \\
&&  \hspace{4cm}-\delta^{[i}_{[l}A_2^{j]}{}_{m]pq}\left(\mathcal{T}^{pq}\right)^{\Lambda}{}_\Omega-A_2^{[j}{}_{lmq}(\mathcal{T}^{i]q})^{\Lambda}{}_\Omega\Big)\\[4pt] \nonumber
&& -12\Big [\tfrac{2}{3}\delta^{[i}_{[l}(\mathcal{T}^{j]r})^{\Lambda}{}_{\Sigma}(\mathcal{T}_{m]r})^{\Sigma}{}_\Omega -\tfrac{1}{2}\delta^{[i}_{[l}(\mathcal{T}_{m]r})^{\Lambda}{}_{\Sigma}(\mathcal{T}^{j]r})^{\Sigma}{}_\Omega \\
&&\nonumber\hspace{1cm} -\tfrac{1}{12}(\mathcal{T}^{ij})^{\Lambda}{}_\Sigma(\mathcal{T}_{lm})^{\Sigma}{}_\Omega+\tfrac{1}{4}(\mathcal{T}_{lm})^{\Lambda}{}_\Sigma(\mathcal{T}^{ij})^{\Sigma}{}_\Omega +\tfrac{1}{4}\delta^{ij}_{lm}(\mathcal{T}_{rs})^{\Lambda}{}_\Sigma(\mathcal{T}^{rs})^{\Sigma}{}_\Omega\Big] \; , \nonumber \\[12pt]
\label{eq:BlockMvec2}
\nonumber \left(\cM^2_\textrm{vec} \right)^{ij \, \Lambda \;lm\Omega}& = &\frac{1}{ 144}A_2^{[i}{}_{qrs}\epsilon^{j]qrsuvw[l}A_2^{m]}{}_{uvw} \delta^{\Lambda\Omega}\\[4pt] \nonumber
&& +\frac{1}{12}\Big(A_2^{[i}{}_{uvw}\epsilon^{j]uvwlmpq}(\mathcal{T}_{pq})^{\Lambda\Omega}-\epsilon^{ijpquvw[l}A_2^{m]}{}_{uvw}(\mathcal{T}_{pq})^{\Lambda\Omega}\Big)\\[4pt] 
&& +12 \Big[\tfrac{1}{4}\Big((\mathcal{T}^{im})^{\Lambda}{}_\Sigma(\mathcal{T}^{jl})^{\Sigma\Omega}-(\mathcal{T}^{jm})^{\Lambda}{}_\Sigma(\mathcal{T}^{il})^{\Sigma\Omega} \\
&&\nonumber \hspace{2cm}+(\mathcal{T}^{il})^{\Lambda}{}_\Sigma(\mathcal{T}^{mj})^{\Sigma\Omega}-(\mathcal{T}^{jl})^{\Lambda}{}_\Sigma(\mathcal{T}^{mi})^{\Sigma\Omega}\Big)\\
&&\nonumber\hspace{2.5cm}-\tfrac{1}{24}(\mathcal{T}^{ij})^{\Lambda}{}_\Sigma(\mathcal{T}^{lm})^{\Sigma\Omega}-\tfrac{1}{48}\epsilon^{ijpqlmrs}(\mathcal{T}_{pq})^{\Lambda}{}_\Sigma(\mathcal{T}_{rs})^{\Sigma\Omega}\Big] \; . \nonumber
\end{eqnarray}
In order to arrive at these expressions, some calculation involving the identities given in appendix B of \cite{Godazgar:2014nqa} is necessary. Against naive expectations, the blocks (\ref{eq:BlockMvec1}), (\ref{eq:BlockMvec2}) cannot be rewritten exclusively in terms of the combined KK fermion shifts $ A_{1 \, i \Lambda , j \Sigma} $ in (\ref{eq:A1KKdd}) above and $A_{2 \, i \Lambda}{}^{jkl \Sigma}$ in (2.26) of \cite{Cesaro:2020soq}. This is reminiscent of the situation for the vector mass matrix, (5.27), (5.28) of \cite{LeDiffon:2011wt}, of $D=4$ $\cN=8$ supergravity with a trombone gauging, which cannot be written either in terms of the relevant fermion shifts solely. Indeed, the ``KK embedding tensor'' (\ref{KKXSymbols}) does bear some resemblance with the trombone embedding tensor of \cite{LeDiffon:2011wt}, along with some crucial differences.

All the eigenvalues of the graviton and gravitino mass matrices, (\ref{eq:KKGravMassMat}), (\ref{eq:A1KKdd}), at a given AdS$_4$ vacuum are physical and correspond to actual spin--2 and spin--$3/2$ KK modes. In contrast, the vector and fermion mass matrices, (\ref{eq:KKVecMassMat}), (\ref{eq:A3dddddd}), contain spurious states at all KK level $n$, corresponding to the magnetic vectors (in the former case), along with Goldstone and Goldstino states eaten by the spin--2 and spin--$3/2$ states. These unphysical states need to be removed, as explained in \cite{Varela:2020wty,Cesaro:2020soq}, from the physical spin-1 and spin--$1/2$ spectra. For all the AdS$_4$ KK spectra covered in \cite{Malek:2020yue,Varela:2020wty,Cesaro:2020soq}, we find experimentally that these vector, $L^2 M^2_{1 \, \textrm{Goldstone}}$, and spin--$1/2$, $L M_{\frac12  \, \textrm{Goldstino}}$, spurious masses, normalised to the radius $L$ of the relevant AdS$_4$ vacuum, are related to graviton, $L^2 M_2^2$, and gravitino, $L M_{\frac32}$, physical masses, at the same KK level and for all levels $n=0, 1 , 2 , \ldots $, through
\begin{equation} \label{eq:Higgsing}
L^2 M^2_{1 \, \textrm{Goldstone}}= 3L^2 M_2^2 +6 \; , \qquad
 L M_{\frac12  \, \textrm{Goldstino}} = 2L M_{\frac32} \; .
\end{equation}
These relations also hold for the KK spectra covered in this paper. We find (\ref{eq:Higgsing}) very helpful to identify the unphysical vector and fermion states to be removed from the spectra. 

The mass matrices (\ref{eq:A1KKdd})--(\ref{eq:KKVecMassMat}) reduce to their counterparts within $D=4$ $\cN=8$ gauged supergravity (see \cite{Trigiante:2016mnt}), and extend those to higher KK levels. Diagonalisation of these four mass matrices is enough to determine the supersymmetric spectrum of the $\cN=1$ AdS$_4$ solutions to any desired KK level. We turn to this in the next section. In all these cases, the scalar dimensions can be deduced from supersymmetry. For this reason, we will not need either to make explicit use of the KK scalar mass matrix \cite{Malek:2020yue}, or to deal with issues related to alternative quantisation \cite{Klebanov:1999tb}.

\section{New spectra of $\cN=1$ AdS$_4$ solutions in $D=11$ and IIA } \label{sec:newspectra}


We now turn to describe the KK spectrum about concrete supersymmetric AdS$_4$ solutions of $D=11$  \cite{Cremmer:1978km}  and massive type IIA \cite{Romans:1985tz} supergravities that uplift consistently on $S^7$ \cite{deWit:1986iy}
and $S^6$ \cite{Guarino:2015jca,Guarino:2015vca} from vacua of the SO(8) \cite{deWit:1982ig} and ISO(7) \cite{Guarino:2015qaa} gaugings of maximal four-dimensional supergravity. See table \ref{tab:AllSusyAdS4} in the introduction for a summary of all known solutions of this type. The complete KK spectrum is now known for all such solutions with $ 2 \leq \cN \leq 8$ supersymmetry \cite{Englert:1983rn,Sezgin:1983ik,Biran:1983iy,Klebanov:2008vq,Malek:2020yue,Varela:2020wty}, and also for the $\cN=1$ solutions with sufficiently large bosonic symmetry groups $G$ \cite{Cesaro:2020soq}. The five AdS$_4$ solutions in this class whose spectrum remains to be given are all $\cN=1$ and display small, if any at all, residual symmetry groups $G$. We have now filled that gap by applying the formalism reviewed in section \ref{sec:KKMassMatRev} to these solutions. In this and the next sections we describe the salient features of these KK spectra, leaving explicit listings for the ancillary files, see appendix \ref{sec:TablesKKSpec}.

Except for the $\textrm{U}(1) \times  \textrm{U}(1)$-invariant vacuum of SO(8)-gauged supergravity \cite{Fischbacher:2009cj,Fischbacher:2010ec} which was reported ten years or so ago, all the other solutions that we will cover here have been discovered fairly recently. These include another vacuum of SO(8)-gauged supergravity with residual SO(3) invariance \cite{Comsa:2019rcz,Bobev:2019dik}; and, in the dyonic ISO(7) gauging, two vacua with U(1) symmetry \cite{Guarino:2019snw} and one more vacuum with no continuous symmetry at all \cite{Bobev:2020qev}. These solutions are only known as critical points of the corresponding $D=4$ $\cN=8$ gauged supergravities, but associated solutions of the form AdS$_4 \times_w S^7$ and AdS$_4 \times_w S^6$ in $D=11$ and massive IIA supergravities are guaranteed to exist by the consistency of the truncation of the latter supergravities down to the former \cite{deWit:1986iy,Guarino:2015jca,Guarino:2015vca}. All these higher dimensional solutions will be warped, supported by internal supergravity forms, and will be equipped with inhomogeneous metrics on the internal spheres with isometry groups containing the residual symmetry groups $G$ of their associated $D=4$ critical points.

As explained in the introduction, the unavailability of the explicit higher-dimensional form of these solutions is not a deterrent to compute their complete KK spectra. All that is needed is the location of the $D=4$ critical points in a given parameterisation of the $\cN=8$ scalar manifold $\textrm{E}_{7(7)}/\textrm{SU}(8)$, {\it i.e.}~the vacuum expectation values (vevs) of the $D=4$ $\cN=8$ scalars at the AdS$_4$ vacua of interest. These vevs have been given in the references above, along with the embedding of the residual symmetry groups into SO(8) or SO(7). These vevs can be brought to the mass matrices (\ref{eq:A1KKdd})--(\ref{eq:KKVecMassMat}), along with the expressions for the SO(8) or SO(7) generators given in appendix A of \cite{Varela:2020wty}, in order to evaluate these mass matrices at each vacuum. Finally, diagonalising these matrices KK level by KK level, and removing the spurious spin--1 and spin--$1/2$ modes discussed at the end of section \ref{sec:KKMassMatRev}, we are able to find the KK towers of physical graviton, gravitino, vector and spin--$1/2$ fermion mass states, $L^2 M_2^2$,  $L M_{\frac32}$, $L^2 M^2_1$, $L M_{\frac12}$, above each of these vacua. It is useful to normalise, as indicated, these masses to the radius of the relevant AdS$_4$ vacuum, $L^2 = -6/V$, with $V<0$ the cosmological constant at the corresponding critical point of the $D=4$ $\cN=8$ supergravity in question. For reference, these cosmological constants can be found in table \ref{tab:AllSusyAdS4} of the introduction. Our conventions for these differ by a factor of 4 with those of the SO(8) survey \cite{Comsa:2019rcz}, but agree with the ISO(7) survey \cite{Bobev:2020qev}.

\newpage

All these AdS$_4$ solutions are $\cN=1$ and preserve a (possibly empty) subgroup $G$ of SO(8) or SO(7) in the $D=11$ or IIA cases, respectively. Accordingly, their KK spectra must organise themselves in representations of $\textrm{OSp}(4|1) \times G$. For all five cases, we have translated the individual KK masses into conformal dimensions via the usual formulae
\begin{equation} \label{eq:LM=Dim}
L^2 M_2^2 = \Delta_2 (\Delta_2 -3) \; , \quad
L^2 M_1^2 = (\Delta_1 -1)  (\Delta_1 -2) \; , \quad
| L M_{\frac32, \frac12} | = \Delta_{\frac32, \frac12} - \tfrac32 \; , 
\end{equation}
(taking always the largest root for the bosonic states) and we have indeed been able to allocate these into OSp$(4|1)$ supermultiplets \cite{Heidenreich:1982rz} KK level by KK level. This is a successful crosscheck of our diagonalisations of the mass matrices (\ref{eq:A1KKdd})--(\ref{eq:KKVecMassMat}). See table 1 of \cite{Cesaro:2020soq} for a summary of the state content of the OSp$(4|1)$ supermultiplets, and for the acronyms with which we will refer to them below. The OSp$(4|1)$ content of the spectra at lowest KK level, $n=0$, is known for all five solutions\footnote{Strictly speaking, for the $D=11$ $\textrm{U}(1) \times \textrm{U}(1)$ solution, the $n=0$ OSp$(4|1)$ spectrum does not seem to have been given in the literature, but it follows from the individual mass states given in \cite{Fischbacher:2009cj,Fischbacher:2010ec}. The $n=0$ bosonic spectrum for the IIA U(1)-invariant solutions was given in \cite{Guarino:2020jwv} and allocated into OSp$(4|1)$ supermultiplets in \cite{Bobev:2020qev}. The OSp$(4|1)$ spectrum for the solutions with SO(3) symmetry in $D=11$ and no continuous symmetry in IIA can be respectively found at KK level $n=0$ in \cite{Bobev:2019dik,Bobev:2020qev}.}. We recover these results and extend them to higher KK levels, $ n \geq 1$. In all cases, we find one and only one massless graviton (MGRAV) multiplet, arising at KK level $n=0$, as expected. Also at level $n=0$, and only at this level, we find a number of massless vector (MVEC) multiplets compatible with the dimension of the residual symmetry group $G$ of each solution: three, two, one or none for the solutions with SO(3), $\textrm{U}(1) \times \textrm{U}(1)$, $\textrm{U}(1)$ or no continuous symmetry. For all the solutions, KK level $n=0$ is completed with a number of massive gravitino (GINO), vector (VEC) and scalar (CHIRAL) multiplets. At every KK level $n \geq 1$, all four generic massive multiplets, GRAV, GINO, VEC and CHIRAL, of OSp$(4|1)$ appear with suitable dimensions $E_0$ for all solutions. As usual, singleton multiplets are absent in all spectra. 

The $D=4$ scalar vevs for all solutions under consideration are only known numerically, except for the 
$\textrm{U}(1) \times \textrm{U}(1)$ solution, where they are known analytically \cite{Fischbacher:2010ec}. Thus, our results for the spectra of all four solutions different than this one are necessarily numerical. For $\textrm{U}(1) \times \textrm{U}(1)$, most of our results are numerical as well, although we have determined analytically some masses and dimensions. Some conformal dimensions stand out as rational or integer within numerical precision. For example, there is a GRAV with $E_0 = \frac92$ in the SO(3) spectrum at level $n=2$. The $\textrm{U}(1) \times \textrm{U}(1)$ spectrum also shows one GINO with $E_0 = 3$, two chirals with $E_0 = 2$ and one CHIRAL with $E_0 = 5$, all of them at $n=2$. The U(1) solution with $g^{-2}c^{\nicefrac13}V=-35.610235$ contains a doubly-degenerate GINO with $E_0 = 3$, and a single GINO with $E_0 = 4$, all of them at KK level $n=1$. Finally, the U(1) solution with $g^{-2}c^{\nicefrac13}V=-25.697101$
has one GINO with $E_0 = 4$ and two GINOs with $E_0=3$ at $n=1$, and two VECs with $E_0 = \frac{11}{2}$ at $n=3$. This list is not exhaustive.

Let us now be more specific about each one of these spectra, particularly regarding degeneracies in them. Degeneracy, or lack thereof, in the conformal dimensions $E_0$ of the generic OSp$(4|1)$ supermultiplets present in the spectra arises in a way compatible with the additional bosonic symmetry $G$ preserved by each solution. Accidental degeneracies also occur for the $G= \textrm{U}(1)$ and $G= \textrm{U}(1) \times \textrm{U}(1)$ invariant solutions, as do for the $\cN=2$ \cite{Malek:2020yue,Varela:2020wty}  and $\cN=3$ \cite{Varela:2020wty}  
solutions of table \ref{tab:AllSusyAdS4}, as well for the $\cN=1$ cases covered in \cite{Cesaro:2020soq}.

\subsubsection*{$D=11$ solution with $\textrm{SO}(3)$ symmetry}

The only degeneracies that appear in the $\cN=1$ spectrum of the $D=11$ SO(3)-invariant solution are those demanded by its SO(3) representation content. In other words, the spectrum arranges itself in $\textrm{OSp}(4|1) \times \textrm{SO}(3)$ representations, with no accidental degeneracies between different representations, either at the same or across different KK levels. This feature singles out this solution together with the $\cN=8$ SO(8) solution, as the only ones in table \ref{tab:AllSusyAdS4} with a continuous residual symmetry and completely non-degenerate supersymmetric spectrum. Except for the type IIA solution with no residual continuous symmetry (which exhibits complete non-degeneracy), the OSp$(4 | \cN)$ spectrum of any other solution with residual continuous symmetry in table \ref{tab:AllSusyAdS4} contains accidental degeneracies.

Another peculiar feature of the KK spectrum of the $\cN=1$ SO(3)-invariant solution is that all the individual states within every OSp$(4|1)$ multiplet have the same charges, not only under SO(3) (as of course they must) but also, somewhat unexpectedly, under a larger $\textrm{SU}(3) \times \textrm{U}(1)_s$. The actual symmetry group SO(3) is embedded into this SU(3) as the real subgroup of the latter (so that the fundamental representation is irreducible), while $\textrm{SU}(3) \times \textrm{U}(1)_s$ is embedded into SO(8) through SO$(7)_s$, with $\bm{8}_s \rightarrow \bm{1}+ \bm{7}$ and $\bm{8}_v$, $\bm{8}_c \rightarrow \bm{8}$ under SO$(7)_s$ so that\footnote{Our conventions are such that, at the $\cN=8$ SO(8) point, the (graviton, gravitini, vectors, spinors, scalars, pseudoscalars) of $\cN=8$ supergravity lie in the $(\bm{1} , \bm{8}_s , \bm{28}  , \bm{56}_s , \bm{35}_v , \bm{35}_c )$ of SO(8). In these conventions, the bosonic symmetries of the $\cN=2$ solutions in table \ref{tab:AllSusyAdS4} are, more precisely, $\textrm{SU}(3) \times \textrm{U}(1)_c$ and $\textrm{SU}(3) \times \textrm{U}(1)_v$ in the SO(8) and ISO(7) gaugings, respectively.}
\begin{equation} \label{eq:SU3U1Branching}
\bm{8}_s \longrightarrow \bm{3}_{-\frac2{3}}+\bar{\bm{3}}_{\frac2{3}}+\bm{1}_{0}+\bm{1}_{0}  \; , \qquad
\bm{8}_v,\; \bm{8}_c \longrightarrow \bm{3}_{\frac1{3}}+\bar{\bm{3}}_{-\frac1{3}}+\bm{1}_{1}+\bm{1}_{-1} 
\end{equation}
under $\textrm{SU}(3) \times \textrm{U}(1)_s$. This is notable for a couple of reasons. Firstly, the symmetry group SO(3) is SO(8)--triality invariant as noted in \cite{Bobev:2019dik}; yet, the KK spectrum shows some preference for the $\bm{8}_s$. Secondly, the spectrum exhibits a qualitative $\textrm{OSp}(4|1) \times\textrm{SU}(3) \times \textrm{U}(1)_s$ structure, even if this group is certainly not a symmetry of the solution and the spectrum does not organise itself in representations of this larger group (because of the SO(3) non-degeneracy just noted). More concretely, the $\textrm{OSp}(4|1) \times\textrm{SO}(3) $ representations in the spectrum branch down from $\textrm{OSp}(4|8)$ via
\begin{equation}
\textrm{OSp}(4|8) \;  \supset  \; \textrm{OSp}(4|1) \times\textrm{SU}(3) \times \textrm{U}(1)_s \;  \supset \; \textrm{OSp}(4|1) \times\textrm{SO}(3) \; ,
\end{equation}
so that, in order to form OSp$(4|1)$ multiplets KK level by KK level, it is enough to split the SO(8) content at each level only under $\textrm{SO}(8) \supset \textrm{SU}(3) \times \textrm{U}(1)_s$.

We are unaware of anything similar happening in the KK spectrum of any other AdS$_4$ solution in table \ref{tab:AllSusyAdS4}. For example, in the KK spectrum of the $\cN=2$  or $\cN=1$ SU(3)--invariant solutions \cite{Klebanov:2008vq,Malek:2020yue,Varela:2020wty,Cesaro:2020soq}, all the individual states within a given OSp$(4|\cN)$ supermultiplet have the same charges under SU(3) (as of course they must). However, different states within the same OSp$(4|\cN)$ multiplet will typically lie in different representations of any larger group containing SU(3), say SU(4) or G$_2$. Of course, this is not surprising, because these larger groups are not symmetries of these solutions. In these cases, one can only form  OSp$(4|\cN)$ multiplets KK level by KK level when the SO(8) in $D=11$ \cite{Englert:1983rn} or  SO(7) in type IIA \cite{Varela:2020wty} state content at each level has already been broken down to the actual residual symmetry group SU(3).

\subsubsection*{$D=11$ solution with $\textrm{U}(1) \times \textrm{U}(1)$ symmetry}

The spectrum of the $D=11$ $\textrm{U}(1) \times \textrm{U}(1)$-invariant  solution displays frequent degeneracies 1, 2 and 4 for the OSp$(4|1)$ dimensions, but also 8 and even 3. The former set of degeneracies, 1, 2 and 4, seems natural for multiplets charged under none, one or both U(1)'s. Any other degeneracy can only be accidental. For example, the  spectrum must contain seven $\textrm{U}(1) \times \textrm{U}(1)$--neutral GRAVs at level $n=2$, as these descend from the $\bm{35}_v$ of SO(8) \cite{Englert:1983rn} and, under $\textrm{SO}(8) \supset \textrm{U}(1) \times \textrm{U}(1)$ \cite{Fischbacher:2010ec},
\begin{eqnarray}
 	\bm{35}_v\; \longrightarrow 7(0,0) + 4 (\pm 1 , 0 ) + 4 (0, \pm 1)  + 3(\pm1,\pm1) + 3(\pm1,\mp1) \; .
\end{eqnarray}
However, there are only two non-degenerate $n=2$ GRAVs, with dimensions $E_0 = \frac{9}{2} = 4.5$ and  $E_0 = 1+ \frac{\sqrt{21}}{2} \approx 3.2912878$. The five remaining singlet GRAV multiplets have dimensions $E_0 = 1+ \frac{\sqrt{37}}{2} \approx 4.0413813$ and $E_0 = 1+ \frac{\sqrt{29}}{2} \approx 3.6925824$ with accidental degeneracies 3 and 2, respectively.

\subsubsection*{Type IIA solutions with U(1) symmetry}

For both type IIA solutions with U(1) symmetry, the dimensions $E_0$ are either non-degenerate or doubly-degenerate. An analysis of the U(1) charges present in both spectra suggests that OSp$(4|1)$ multiplets with non-vanishing, opposite U(1) charges are always degenerate. These appear as doubly-degenerate multiplets in the tables. All non-degenerate multiplets are in turn U(1)--neutral. The converse is not true, however: some U(1)--neutral multiplets are accidentally doubly-degenerate.

In order to see this, let us look for example at the spectrum of GRAV and GINO multiplets at KK level $n=1$. The individual spin--2 and spin--$3/2$ states contained therein have U(1) charges that respectively descend from the representations $\bm{7}$ and $\bm{8}+\bm{48}$ of SO(7) \cite{Varela:2020wty}. Under the embedding $\textrm{SO}(7) \supset \textrm{U}(1)$ described in \cite{Guarino:2019snw},
\begin{equation}\label{U(1)branchings}
\bm{7} \rightarrow  3(0) + 2 \left(\pm \tfrac12 \right) \; , \quad
\bm{8} \rightarrow  4(0) + 2 \left(\pm \tfrac12 \right) \; , \quad
\bm{48} \rightarrow  16 (0) + 12 \left(\pm \tfrac12 \right) + 4 ( \pm 1)  \; ,
\end{equation}
in line with the 3 non-degenerate and 2 doubly degenerate $n=1$ GRAVs shown in either spectrum. Each of these also shows 12 non-degenerate and 15 doubly-degenerate GINOs at level $n=1$. The branchings (\ref{U(1)branchings}) are compatible with all 12 non-degenerate GINOs being U(1)-neutral, 14 doubly-degenerate GINOs being charged, and two further U(1)-neutral GINO multiplets being accidentally degenerate.

\subsubsection*{Type IIA solution with no continuous symmetry}

The KK spectrum of the type IIA solution with no continuous symmetry is completely non-degenerate. Indeed, the conformal dimension $E_0$ of every single OSp$(4|1)$ multiplet present in the spectrum is different. This spectrum thus plays by the book, making no concessions whatsoever to accidental degeneracies.

\vspace{10pt}

We conclude by emphasising that, by $\cN=1$ supersymmetry, our results also contain the KK scalar spectrum above all these solutions, even if we did not explicitly diagonalise the KK scalar mass matrix. Please refer to the next section for further comments.

\section{Further discussion} \label{sec:discussion}


The $\cN=1$ KK spectra that we have covered in this paper exhibit some similarities and some differences with respect to the spectra previously computed for the other solutions in table \ref{tab:AllSusyAdS4} of the introduction. For example, the $\textrm{U}(1) \times \textrm{U}(1)$- and U(1)-invariant solutions are like other previously studied cases in that accidental degeneracies occur in their spectra. Also, these solutions enjoy discrete symmetries, but their potential effect on the spectra is unclear. Here, we will discuss another feature in which all five new spectra in this paper differ from the other cases previously studied: the absence of an overarching formula for the conformal dimensions of the multiplets present in the spectra --or, at least, a formula of the same type than that for the previous cases.

A single master formula \cite{Cesaro:2020soq},
\begin{equation} \label{eq:E0genericN}
 \quad E_0 =  s_0^\2 -\tfrac12 + \sqrt{ \tfrac{9}{4} + s_0^\2 (s_0^\2 +1) -s_0(s_0+1) + \alpha \, n (n + d-1) + {\cal Q}_2 (R) }  \; ,
\end{equation}
governs the conformal dimension $E_0$ of {\it any} OSp$(4|\cN)$ supermultiplet with conformal primary spin $s_0$, present in the spectrum at KK level $n$ and in a representation $R$ of the residual symmetry group $G$, for all of the $D=11$ ($d=7$) or type IIA ($d=6$) AdS$_4$ solutions in table \ref{tab:AllSusyAdS4} with supersymmetry $2 \leq \cN \leq 8$. The same formula also applies for the $\cN=1$ solutions in the table with sufficiently large bosonic symmetry groups, $G = \textrm{G}_2$ or $G = \textrm{SU}(3)$. In (\ref{eq:E0genericN}), $s_0^\2$ and $\alpha$ are, respectively, half-integer and rational constants, different for each solution but independent of $n$ and $R$. Also, ${\cal Q}_2$ is a quadratic, invariant polynomial on the $G$ quantum numbers of the supermultiplet, {\it i.e.} on the Dynkin labels characterising $R$. The only dependence of the r.h.s.~of (\ref{eq:E0genericN}) on the KK level $n$ is, directly, through the explicit contribution $ n (n + d-1) $ and, indirectly, through ${\cal Q}_2$, as the quantum numbers are typically bounded by $n$. Please refer to \cite{Cesaro:2020soq} for full details.

That the KK spectrum of the $D=11$ SO(3)-invariant solution cannot abide by the master formula (\ref{eq:E0genericN}) may be argued from the lack of accidental degeneracies noted in section \ref{sec:newspectra}. This formula is independent of the number of times a given supermultiplet arises in the spectrum at fixed KK level $n$ and in the same representation $R$. The spectrum for the SO(3) solution contains many such repeated multiplets, none of them though with the same dimension $E_0$. For example, there are two triplet and two singlet GRAVs at KK level $n=1$, compatible with the fact that
\begin{eqnarray} \label{eq:8vunderSO3}
\bm{8}_v \longrightarrow \bm{3} +\bm{3} +\bm{1} +\bm{1} \; ,
\end{eqnarray}
under $\textrm{SO}(8) \supset \textrm{SO}(3)$. Formula (\ref{eq:E0genericN}) with $s_0^\2 = s_0 = \frac32$, some fixed $\alpha$ and ${\cal Q}_2 \sim \ell (\ell+1)$, where $\ell$ is the SO(3) spin, predicts the same dimension for both triplets and both singlets, something that according to our data does not happen. One may try and generalise (\ref{eq:E0genericN}) by, for example, letting ${\cal Q}_2$ depend more generally on $\ell$ and on the additional quantum numbers of the $\textrm{SU}(3) \times \textrm{U}(1)_s$ discussed in section \ref{sec:newspectra}, akin to the Zeeman effect in the Hydrogen atom. At least for the $n=1$ GRAVs, this has chances to do the trick because the SO(3) representations in (\ref{eq:8vunderSO3}) are distinguished by the charges under the broken U$(1)_s$: see (\ref{eq:SU3U1Branching}). Nevertheless, this does not work in general: other multiplets in the KK spectrum have the same $\textrm{SU}(3) \times \textrm{U}(1)_s$ quantum numbers and still non-degenerate dimensions $E_0$. 

With its complete non-degeneracy, the type IIA solution with no continuous symmetry deviates even more dramatically from (\ref{eq:E0genericN}). The $\textrm{U}(1) \times \textrm{U}(1)$- and the $\textrm{U}(1)$-invariant solutions do show some degeneracy, but still their dimensions do not obey a formula like (\ref{eq:E0genericN}) either. The reason why the dimension formula (\ref{eq:E0genericN}) works, at least for the spectrum of GRAV multiplets above all the solutions covered previously, may be traced back to the fact that, in those cases, the characteristic polynomials of the graviton mass matrix (\ref{eq:KKGravMassMat}) at every KK level $n$ are always products of polynomials, irreducible over the rational numbers, of at most degree two. For example, for the $D=11$ $\mathcal{N}=1$ G$_2$-invariant solution, the characteristic polynomial of the graviton mass matrix at KK level $n=3$ is
\begin{equation}
\mathcal{CP}^{\text{G}_2}_{n=3}(\lambda) =  \left(\lambda-\tfrac{135}{8}\right)\left(\lambda-\tfrac{115}{8}\right)^7( \lambda-\tfrac{265}{24})^{27}( \lambda-\tfrac{55}{8})^{77} \left(\lambda-\tfrac{35}{8}\right)\left(\lambda-\tfrac{15}{8}\right)^7\; .
\end{equation}
In contrast, the characteristic polynomials for the mass matrices at the $\textrm{U}(1) \times \textrm{U}(1)$ point do not follow this pattern. Instead, polynomial factors appear in them that are irreducible over the rational numbers and have degrees that increase with the KK level $n$. For example, for the $n=1$ gravitons one finds a quadratic polynomial
\begin{equation} \label{eq:CharPolU(1)Sqn=1}
 	\mathcal{CP}^{\text{U(1)}^2}_{n=1}(\lambda)=\left(16 \lambda^2-104\lambda+121\right)^4\;,
\end{equation}
repeated four times per the relevant multiplicity. This is similar to the situation for the previously studied cases, except that the quadratic in (\ref{eq:CharPolU(1)Sqn=1}) is already irreducible over the rationals. At KK level $n=2$ a rational-irreducible quadratic again appears and, more importantly, there is also a cubic:
\begin{eqnarray}	\label{eq: chpolyn2}
	 \mathcal{CP}^{\text{U(1)}^2}_{n=2}(\lambda)\!&=&\!\left(\lambda-10\right)\big(\lambda-8\big)^4\big(\lambda-\tfrac{29}4\big)^4\big(\lambda-7\big)^3\big(\lambda-5\big)^2\left(\lambda-3\right)\\ \nonumber
                                &&\hspace{2cm}\quad \times\big(\lambda^2-17\lambda+45\big)^4\big( 64 \lambda^3- 1392 \lambda^2 + 8684 \lambda-15429\big)^4    \; .
\end{eqnarray}
At level $n=3$, quartic, sextic and rational-irreducible factors of degree 12 appear:
{\footnotesize
\begin{eqnarray}	\label{eq: chpolyn3}
	\mathcal{CP}^{\text{U(1)}^2}_{n=3}(\lambda)\!&=&\!\big(256 \lambda^4-15616 \lambda^3+325984 \lambda^2-2703696 \lambda+7295409\big)^4\notag\\[4pt]
			&&\times\big(4096 \lambda^6-321536 \lambda^5+10067712 \lambda^4-160656128 \lambda^3+1373650672 \lambda^2\notag\\[4pt]
			&&\quad-5939292648 \lambda+10065721617\big)^8		\\[4pt]
			&&\times\big(16777216 \lambda^{12}-2231369728 \lambda^{11}+134395985920 \lambda^{10}-4846825504768 \lambda^9	\notag\\[4pt]
			&&\quad+116551324205056 \lambda^8-1968472046829568 \lambda^7+23938429294919680 \lambda^6		\notag\\[4pt]
			&&\qquad-211148817120108544 \lambda^5+1340283207107845888 \lambda^4-5968527029434695936 \lambda^3		\notag\\[4pt]
			&&\quad\qquad+17691501531337880736 \lambda^2-31321971689759456400 \lambda+25030390172750615673\big)^4\; . \notag
\end{eqnarray}
}All these lead to graviton mass eigenvalues which are not rational or square roots of rational numbers. None of these have dimensions, associated via (\ref{eq:LM=Dim}), compatible with (\ref{eq:E0genericN}). A qualitatively similar argument can also be made for all other four solutions covered in this paper, although in those cases the polynomials have approximate numerical coefficients. 

\begin{figure}
\centering
	\begin{subfigure}{.42\textwidth}
		\centering
		\includegraphics[width=1.0\linewidth]{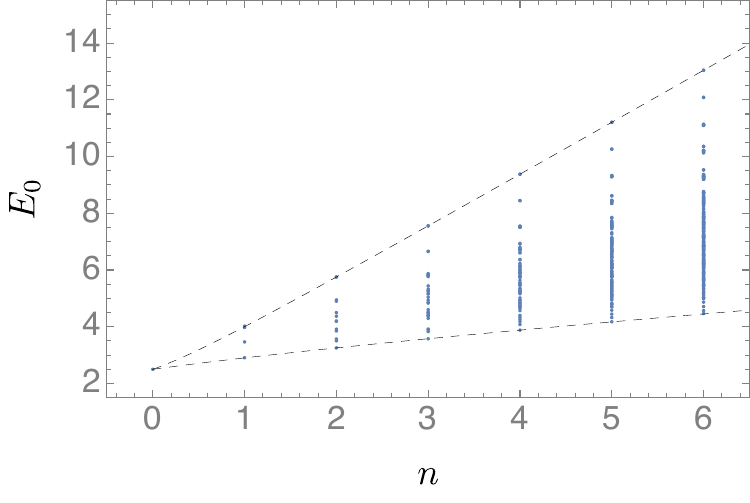}
  		\caption{\footnotesize{(M)GRAVs}}
		\label{fig: gravsso3}
	\end{subfigure}%
	\quad
	\begin{subfigure}{.42\textwidth}
		\centering
		\includegraphics[width=1.0\linewidth]{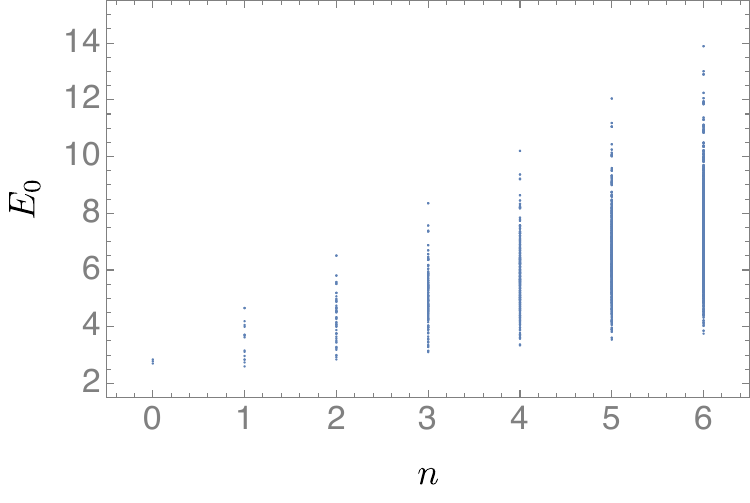}
  		\caption{\footnotesize{GINOs}}
		\label{fig: ginoso3}
	\end{subfigure}\\[8pt]
	\begin{subfigure}{.42\textwidth}
		\centering
		\includegraphics[width=1.0\linewidth]{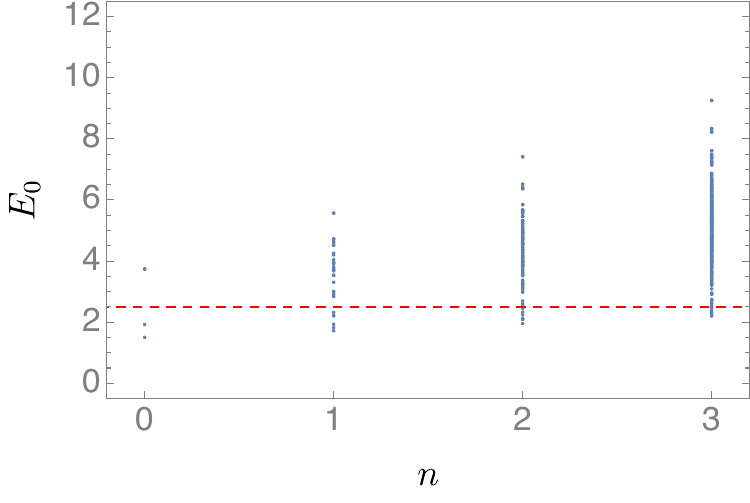}
  		\caption{\footnotesize{(M)VECs}}
		\label{fig: vecso3}
	\end{subfigure}%
	\quad
	\begin{subfigure}{.42\textwidth}
		\centering
		\includegraphics[width=1.0\linewidth]{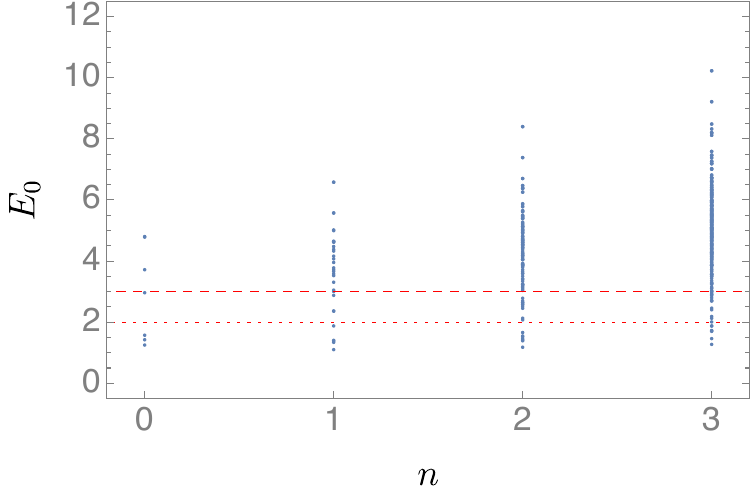}
  		\caption{\footnotesize{CHIRALs}}
		\label{fig: chiralso3}
	\end{subfigure}
	\caption{\footnotesize{Dimensions $E_0$ of all four types of OSp$(4|1)$ multiplets present in the  spectrum of the $D=11$ SO(3)-invariant solution at first few KK levels $n$.} \label{fig:so3sol} }
\end{figure}

Given the lack of a formula like (\ref{eq:E0genericN}) for our five new $\cN=1$ spectra, it is helpful to resort to plots to visualise the data contained in the ancillary files and find qualitative trends. Two such representative graphs have been taken to figures \ref{fig:so3sol} and \ref{fig:u12sol}. These respectively plot all the dimensions $E_0 = E_0(n)$, for the first few KK levels $n$ and for all types of OSp$(4|1)$ multiplets, contained in the spectra of the $D=11$ SO(3)-invariant solution and the type IIA solution with U(1) invariance and cosmological constant $g^{-2}c^{\nicefrac13}V=-35.610235$. As we will discuss below, these plots reveal some patterns consistent with those observed in related non-supersymmetric contexts \cite{Malek:2020mlk,Guarino:2020flh}, now extrapolated to an $\cN = 1$ setting. They also contain some little surprises.

For each type of multiplet present at KK level $n$, it is useful to look at the highest, $E_0^{\text{max}}(n)$, and the lowest, $E_0^{\text{min}}(n)$, dimension among all multiplets of the same type at that KK level. For all four generic types of multiplets in all five spectra addressed in this paper, these maximal dimensions $E_0^{\text{max}}(n)$ exhibit a monotonically growing trend with the KK level $n$: $E_0^{\text{max}}(n_1) < E_0^{\text{max}}(n_2)$ for $n_1 < n_2$. For example, for the SO(3) and U(1), $g^{-2}c^{\nicefrac13}V=-35.610235$ solutions, this is apparent from figures \ref{fig:so3sol} and \ref{fig:u12sol}. The same growing trend of  $E_0^{\text{max}}(n)$ is also present for all supermultiplets in the three remaining $\cN=1$ solutions in table \ref{tab:AllSusyAdS4} of the introduction, whose spectra were covered in \cite{Cesaro:2020soq}. In contrast, the minimal dimensions $E_0^{\text{min}}(n)$ show two types of qualitative behaviour: either a monotonic increase as well, or non-monotonic behaviour in the first few KK levels finally followed by growth, at least on average, with $n$. We will refer to these two cases as monotonic and non-monotonic, respectively. Table \ref{tab:monotonicbehaviour} below summarises the behaviour of $E_0^{\text{min}}(n)$ in this regard, for each OSp$(4|1)$ multiplet in the spectra of all eight $\cN=1$ solutions of table \ref{tab:AllSusyAdS4}.

\begin{figure}
\centering
	\begin{subfigure}{.42\textwidth}
		\centering
		\includegraphics[width=1.0\linewidth]{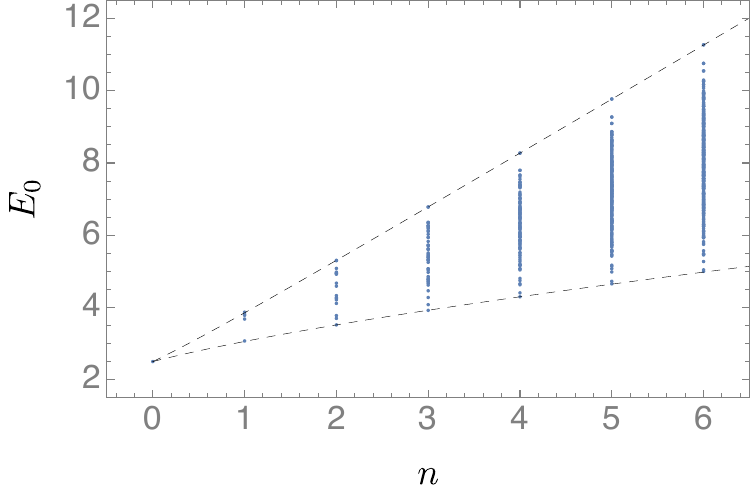}
  		\caption{(M)GRAVs}
		\label{fig: gravsu12}
	\end{subfigure}%
	\quad
	\begin{subfigure}{.42\textwidth}
		\centering
		\includegraphics[width=1.0\linewidth]{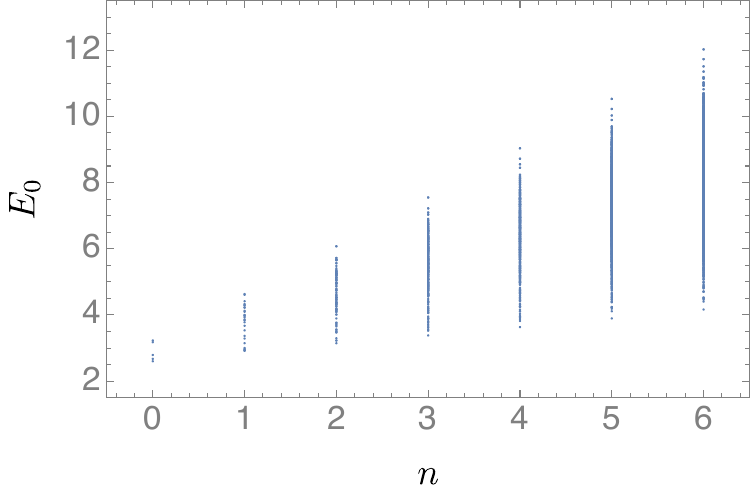}
  		\caption{GINOs}
		\label{fig: ginou12}
	\end{subfigure}\\[8pt]
	\begin{subfigure}{.42\textwidth}
		\centering
		\includegraphics[width=1.0\linewidth]{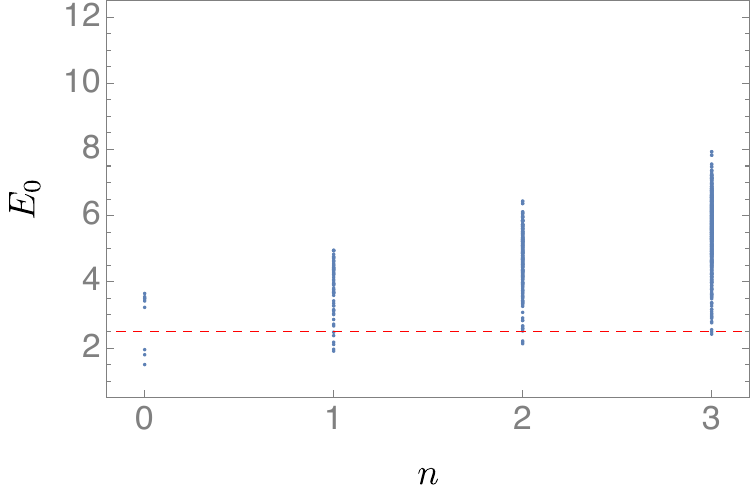}
  		\caption{(M)VECs}
		\label{fig: vecu12}
	\end{subfigure}%
	\quad
	\begin{subfigure}{.42\textwidth}
		\centering
		\includegraphics[width=1.0\linewidth]{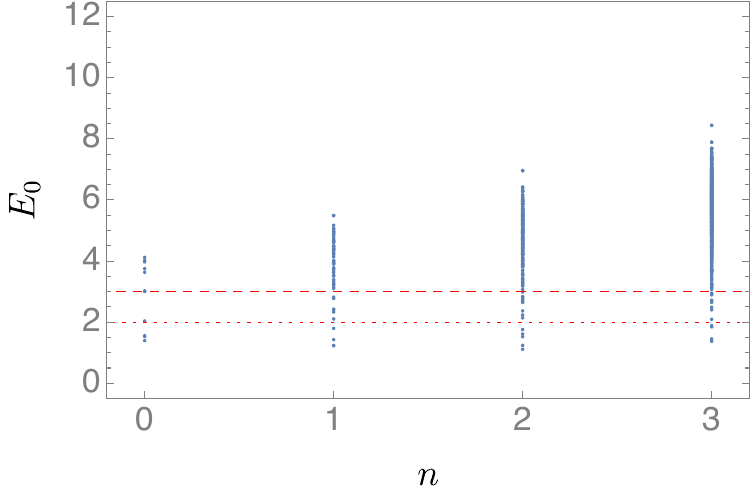}
  		\caption{CHIRALs}
		\label{fig: chiralu12}
	\end{subfigure}
	\caption{\footnotesize{Dimensions $E_0$ of all four types of OSp$(4|1)$ multiplets present in the  spectrum of the type IIA U(1)-invariant solution with cosmological constant  $g^{-2}c^{\nicefrac13}V=-35.610235$, at first few KK levels $n$.} \label{fig:u12sol} }
\end{figure}

In all eight $\cN=1$ spectra, $E_0^{\text{min}}(n)$ for the (M)GRAV and the (M)VEC supermultiplets is monotonic. For the SO(3) and the U(1), $g^{-2}c^{\nicefrac13}V=-35.610235$ spectra, for example, this can be seen from figures \ref{fig: gravsso3}, \ref{fig: vecso3}, \ref{fig: gravsu12} and \ref{fig: vecu12}. The minimal dimensions for the GINO multiplets are also monotonic, except for the SO(3) solution, for which $E_0^{\text{min}}(n)$ exhibits non-monotonicity. This can be seen in figure \ref{fig: ginoso3}, where $E_0^{\text{min}}(0) > E_0^{\text{min}}(1)$, only to bounce back at $n=1$ and keep growing thereafter. The non-monotonic behaviour of $E_0^{\text{min}}(n)$ is even more present for the CHIRAL multiplets. For the $D=11$ solutions, $E_0^{\text{min}}(n)$ for these multiplets is monotonic for the G$_2$ spectrum, but non-monotonic for the $\textrm{U}(1) \times \textrm{U}(1)$ and SO(3) spectra. For the latter case this can be seen from figure \ref{fig: chiralso3}, where $E_0^{\text{min}}(0) > E_0^{\text{min}}(1)$ but then $E_0^{\text{min}}(1) < E_0^{\text{min}}(2)$. The non-monotonic behaviour of the scalar multiplets in these two $D=11$ cases is qualitatively similar to that observed in \cite{Malek:2020mlk} for the KK scalar masses above the non-supersymmetric $D=11$ solution of \cite{Warner:1983du,Godazgar:2014eza}. Of course, in the present $\cN=1$ cases, the bounce is always above the Breitenlohner-Freedman bound for stability on AdS$_4$, unlike in \cite{Malek:2020mlk}. In the type IIA cases, $E_0^{\text{min}}(n)$ shows monotonic behaviour for the CHIRALs in the spectra of the solutions with G$_2$, SU(3) and U(1) symmetry with $g^{-2}c^{\frac13}V=-25.697101$. These solutions thus obey an $\cN=1$ version of the monotonic behaviour shown in \cite{Guarino:2020flh} to hold for the minimal scalar masses at each KK level in the spectra of the non-supersymmetric AdS$_4$ solutions covered therein. However,  $E_0^{\text{min}}(n)$ is non-monotonic for the IIA solutions with no continuous symmetry and U(1) symmetry with $g^{-2}c^{\nicefrac13}V=-35.610235$, as seen for the latter in figure \ref{fig: chiralu12}. This perhaps comes as a slight surprise given the results of \cite{Guarino:2020flh}, and  makes more plausible the potential existence of non-supersymmetric vacua in the ISO(7) gauging with perturbative instabilities only at some higher KK level $n \geq 1$ but not at $n=0$, like in \cite{Warner:1983du,Godazgar:2014eza,Malek:2020mlk} for the SO(8) gauging.

\begin{table}
%
\resizebox{\textwidth}{!}{
\begin{tabular}{ccccccccc}
\hline&&&\\[-2.5mm]
\multirow{1}{*}{OSp(4$\vert$1) mult.}  	& G$_2$ &  SO(3) &  U(1)$\times$U(1) & G$_2$ & SU(3) & U(1) & U(1) & $\emptyset$	\\[0pt]
				        	                         &  {\footnotesize in $D=11$}           &             &   &   {\footnotesize in type IIA}           &           & {\footnotesize $g^{-2}c^{\frac13}V=-25.697$} & {\footnotesize $g^{-2}c^{\frac13}V=-35.610$} &	\\[2pt]
\hline&&&\\[-10pt] 
(M)GRAV	&      \mon       &  \mon   &  \mon & \mon & \mon &  \mon &  \mon &  \mon   \\[10pt]
GINO	&      \mon       &  \nonmon   &  \mon  & \mon & \mon &  \mon &  \mon &  \mon \\[10pt]
(M)VEC		&      \mon       &  \mon   &  \mon  & \mon & \mon &  \mon &  \mon &  \mon \\[10pt]
CHIRAL	&       \mon     &  \nonmon   &  \nonmon    & \mon & \mon &  \mon &  \nonmon &  \nonmon \\[5pt]
\hline
\end{tabular}
}

\caption[]{\footnotesize{Monotonicity (\resizebox{!}{0.5\baselineskip}{\!\mon}) or non-monotonicity (\resizebox{!}{0.8\baselineskip}{\!\!\!\!\!\nonmon\!\!\!\!}) of $E_0^{\text{min}}(n)$ for all OSp$(4|1)$ multiplets in the KK spectra of all $\cN=1$ solutions of table \ref{tab:AllSusyAdS4}.
}\normalsize}
\label{tab:monotonicbehaviour}
\end{table}


Even if a unique formula similar to (\ref{eq:E0genericN}) does not exist for the dimension of every single OSp$(4|1)$ supermultiplet in our new $\cN=1$ spectra, one may still enquire whether $E_0^{\text{max}}(n)$ and $E_0^{\text{min}}(n)$ conform to some simple mathematical expression. It seems reasonable to assume that these extremal dimensions correspond to limiting values of the quantum numbers of the residual symmetry group $G$ at each solution. As mentioned above, these are typically bounded by $n$. Maintaining a square-root type of formula like (\ref{eq:E0genericN}) with radicand quadratic in the quantum numbers, we are led to consider the ansatz:
\begin{equation}	\label{eq: bounds}
	E_0^{\text{max}}(n) =1+\sqrt{ a^{\text{max}} + b^{\text{max}}  \, n + c^{\text{max}} \,  n^2} \;  ,
\end{equation}
and similarly for $E_0^{\text{min}}(n)$. Here, $a^{\text{max}}$, $b^{\text{max}}$ and $c^{\text{max}}$ are constants, different for each type of multiplet and for each solution. The proposal (\ref{eq: bounds}) turns out to fit very well the actual $E_0^{\text{max}}(n)$ and $E_0^{\text{min}}(n)$ for the GRAVs across all five spectra, with $a^{\text{max}} = a^{\text{min}} = \tfrac94$, as in the cases discussed in \cite{Cesaro:2020soq}, and appropriate values for the other constants. The fit (\ref{eq: bounds}) is depicted for the SO(3) and U(1), $g^{-2}c^{\nicefrac13}V=-35.610235$ GRAVs with dashed lines in figures \ref{fig: gravsso3} and \ref{fig: gravsu12}. For other supermultiplets, (\ref{eq: bounds}) yields mixed results. This is not surprising: simple anayltical formulae for the extremal dimensions tend not to exist even in the cases considered in \cite{Cesaro:2020soq}, for which the individual dimensions (\ref{eq:E0genericN}) are known.

One may finally enquire about the presence of relevant ($\Delta <3$) or classically marginal ($\Delta = 3$) scalar states in our KK spectra. Our data up to KK level $n =3$ for the supermultiplets, VEC and CHIRAL, that contain scalars do not allow for a conclusive classification at all KK level, especially for the spectra in which $E_0^{\text{min}}(n)$ for these multiplets is non-monotonic. With our data and across all five spectra, there is only one multiplet containing $\Delta = 3$ scalars. It is an $E_0 = 2$ CHIRAL multiplet in the $\textrm{U}(1) \times \textrm{U}(1)$ spectrum at level $n=2$. In figures \ref{fig:so3sol} and \ref{fig:u12sol}, a visual reference of multiplets that contain $\Delta \leq 3$ scalars within the SO(3) and U(1), $g^{-2}c^{\nicefrac13}V=-35.610235$ spectra is given. Specifically, in figures \ref{fig: vecso3} and \ref{fig: vecu12}, VEC multiplets with dimensions greater than, equal to, or lower than $E_0 = \tfrac52$ (corresponding to the horizontal dashed red line), respectively contain one irrelevant, classically marginal, or relevant scalar. This is similar for the CHIRALs in figures \ref{fig: chiralso3} and \ref{fig: chiralu12}, with the red dashed line now sitting at $E_0=3$. In the latter figures, CHIRALs with $E_0 < 2$ (marked by the red dotted line), further contain a second relevant scalar.


\section*{Acknowledgements}


We would like to thank Emilio Ambite, \'Alvaro Mu\~noz, Carlos Pena, Salvador Rosauro and Miguel Sacrist\'an for useful comments on cluster computing and numerical methods. The numerical calculations of this paper were performed on the Hydra cluster of IFT. MC is supported by a La Caixa Foundation (ID 100010434) predoctoral fellowship LCF/ BQ/DI19/11730027. GL is supported by an FPI-UAM predoctoral fellowship. OV is supported by the NSF grant PHY-2014163. MC, GL and OV are partially sup\-por\-ted by grants SEV-2016-0597 and PGC2018-095976-B-C21 from MCIU/AEI/FEDER, UE.

A previous version of this paper contained tabulated listings of the supersymmetric spectra where some spurious modes had leaked in. The tables have been removed in the present version, and the correct data is now presented in ancillary files. We thank Nikolay Bobev for pointing out these typos.


\appendix

\section{Ancillary files} \label{sec:TablesKKSpec}

The files attached to this submission contain our results for the supersymmetric KK spectra corresponding to the five $\cN=1$ AdS$_4$ solutions of $D=11$ or massive IIA supergravities that we have covered in this paper. For each solution, SO(3), $\textrm{U}(1) \times \textrm{U}(1)$, U(1) with $g^{-2}c^{\nicefrac13}V=-25.697101$, U(1) with $g^{-2}c^{\nicefrac13}V=-35.610235$ and the solution with no continuous symmetry,  there is a file, \texttt{supermultiplets\_SO3.wl}, \texttt{supermultiplets\_U1U1.wl}, \texttt{supermultiplets\_U1\_V2569.wl}, \texttt{supermultiplets\_U1\_V3561.wl} and
\texttt{supermultiplets} \texttt{\_nosymm.wl}. This type, \texttt{.wl}, of file is simply a plain text file, which can thus be opened with any text editor, and which Wolfram Mathematica can load to memory with the \texttt{Get} command. Each of these files gives the multiplicities and the dimensions of the OSp$(4|1)$ supermultiplets, MGRAV, GRAV, GINO, MVEC, VEC and CHIRAL, present in the spectrum for that solution up to KK level $n=3$. This data arises by appropriately combining the eigenvalues of the mass matrices (\ref{eq:A1KKdd})--(\ref{eq:KKVecMassMat}), translated to dimensions via (\ref{eq:LM=Dim}). The fact that these eigenvalues successfully combine into supermultiplets provides a consistency check on our diagonalisations. All the OSp$(4|1)$ multiplets present in these tables contain only physical states: prior to allocation into multiplets, all spurious states have been removed as indicated at the end of section \ref{sec:KKMassMatRev}. See {\it e.g.}~table 1 of \cite{Cesaro:2020soq} for the state content of these multiplets.

The data are presented in lists that have been given intuitive names. Each of these lists contains all multiplets of a given type present in the spectrum at the indicated KK level. For example, the file \texttt{supermultiplets\_SO3.wl} contains the list
\begin{equation}
\textrm{ \texttt{ E0MGRAVkk0 = \{\{2.5, 1\}\} }} \; ,
\end{equation}
which informs us of the presence in the spectrum of the SO(3) solution of a single MGRAV with dimension $E_0 = 2.5$  and multiplicity one at KK level $n=0$. Similarly, in the same file, the variable
\begin{equation}
\textrm{ \texttt{ E0GINOkk1 = \{\{4.661952018343944, 3\}, \{4.656076036620153, 3\}, 
 $\ldots$ \}  }} \; ,
\end{equation}
lists all GINO multiplets at KK level $n=1$: there are three such multiplets with dimension $E_0 \approx 4.66195$, three multiplets with dimension $E_0 \approx 4.65608$, etc. We have presented the results with fifteen decimal places.

\bibliography{references}

\end{document}